\documentclass[10pt]{article}

\begin{document}

\newcommand{\df}{\stackrel{\rm def}{=}}
\newcommand{\co}{{\scriptstyle \circ}}
\newcommand{\lb}{\lbrack}
\newcommand{\rb}{\rbrack}
\newcommand{\rn}[1]{\romannumeral #1}
\newcommand{\msc}[1]{\mbox{\scriptsize #1}}
\newcommand{\dsp}{\displaystyle}
\newcommand{\scs}[1]{{\scriptstyle #1}}

\newcommand{\ket}[1]{| #1 \rangle}
\newcommand{\bra}[1]{| #1 \langle}
\newcommand{\vac}{| \mbox{vac} \rangle }

\newcommand{\e}{\mbox{{\bf e}}}
\newcommand{\va}{\mbox{{\bf a}}}
\newcommand{\bc}{\mbox{{\bf C}}}

\newcommand{\com}{C\!\!\!\!|}

\newcommand{\br}{\mbox{{\bf R}}}
\newcommand{\bz}{\mbox{{\bf Z}}}
\newcommand{\bq}{\mbox{{\bf Q}}}
\newcommand{\bn}{\mbox{{\bf N}}}
\newcommand {\eqn}[1]{(\ref{#1})}

\newcommand{\cp}{\mbox{{\bf P}}^1}
\newcommand{\n}{\mbox{{\bf n}}}
\newcommand{\sbz}{\msc{{\bf Z}}}
\newcommand{\sn}{\msc{{\bf n}}}

\newcommand{\be}{\begin{equation}}\newcommand{\ee}{\end{equation}}
\newcommand{\bea}{\begin{eqnarray}} \newcommand{\eea}{\end{eqnarray}}
\newcommand{\ba}[1]{\begin{array}{#1}} \newcommand{\ea}{\end{array}}

\newcommand{\cleqn}{\setcounter{equation}{0}}

\makeatletter

\@addtoreset{equation}{section}

\def\theequation{\thesection.\arabic{equation}}
\makeatother

\def\np{Nucl. Phys. {\bf B}}\def\pl{Phys. Lett. {\bf B}}
\def\mpl{Mod. Phys. {\bf A}}\def\ijmp{Int. J. Mod. Phys. {\bf A}}
\def\cmp{Comm. Math. Phys.}\def\prd{Phys. Rev. {\bf D}}

\def\oa{\bigcirc\!\!\!\! a}
\def\ob{\bigcirc\!\!\!\! b}
\def\oc{\bigcirc\!\!\!\! c}
\def\oi{\bigcirc\!\!\!\! i}
\def\oj{\bigcirc\!\!\!\! j}
\def\ok{\bigcirc\!\!\!\! k}
\def\ve{\vec e}\def\vk{\vec k}\def\vn{\vec n}\def\vp{\vec p}
\def\vr{\vec r}\def\vs{\vec s}\def\vt{\vec t}\def\vu{\vec u}
\def\vv{\vec v}\def\vx{\vec x}\def\vy{\vec y}\def\vz{\vec z}

\def\ve{\vec e}\def\vk{\vec k}\def\vn{\vec n}\def\vp{\vec p}
\def\vr{\vec r}\def\vs{\vec s}\def\vt{\vec t}\def\vu{\vec u}
\def\vv{\vec v}\def\vx{\vec x}\def\vy{\vec y}\def\vz{\vec z}


\begin{flushright}
La Plata Th/07-01\\May, 2007
\end{flushright}

\bigskip

\begin{center}

{\Large\bf On supersymmetric $D6$-$\bar D 6$ systems\\with magnetic fields}
\footnote{
This work was partially supported by CONICET, Argentina
}
\bigskip
\bigskip

{\it \large Walter H. Baron and Adri\'{a}n R. Lugo}
\footnote{
{\sf w$\_$baron@iafe.uba.ar, lugo@fisica.unlp.edu.ar}
}
\bigskip

{\it Departamento de F\'\i sica and IFLP-CONICET\\
Facultad de Ciencias Exactas , Universidad Nacional de La Plata
\\ C.C. 67, (1900) La Plata, Argentina}
\bigskip
\bigskip

\end{center}
\bigskip

\begin{abstract}
We study systems of  $D6$ and $\bar D 6$ branes with non zero world-volume
magnetic fields in the weak coupling limit.
We find two configurations for which the conditions for absence of tachyons in the spectra coincide exactly with
those found in the low energy effective theory approach, for the systems
to preserve $\frac{1}{8}$ of the supersymmetries of the Type $IIA$ string theory vacuum.
These conditions give rise to a four-parameter family of solutions in each case.
We present further evidence of the stability of these systems by computing the lowest order
interaction amplitude, verifying the no force condition as well as the supersymmetric character of the spectrum.
\end{abstract}
\bigskip

\section{Introduction}
\cleqn

Some years ago, supersymmetric (SUSY) cylinder-like brane configurations of arbitrary section,
with $D0$ and $F1$ charges and no $D2$ charge, the so-called ``supertubes", were discovered \cite{Mateos:2001qs}, \cite{Emparan:2001ux}
(see also \cite{Bak:2006yb}).
They can be seen as true bound states of fundamental strings and $D0$ branes.
Furthermore, the supertubes preserve one quarter of the supersymmetries of the flat Minkowski space-time vacuum of type IIA string theories,
the stabilizing factor at the origin of their BPS character (which prevents them from collapsing) being the angular momentum generated by
the non-zero electric and magnetic fields that live on the brane.
This discovery leads  to conjecture the existence of SUSY $D2-\bar D2$ systems.
The argument is simple: If we consider, for example, a supertube of elliptic section in the limit when one of the semi-axis  goes to infinity,
the resulting system should be equivalent to having two flat $2$-branes with total $D2$ charge equal to zero.
The study of systems with arbitrary numbers of $D2$ and $\bar D2$ branes was made in the context of the Dirac-Born-Infeld (DBI) action in reference
\cite{Bak:2001xx}, where the conditions to be satisfied by the Killing spinors were identified.
Soon after that, in references \cite{Bak:2002wy}, \cite{Lugo:2002wr}, higher dimensional brane-antibrane systems were
considered in the DBI context, and the existence of one quarter SUSY $D4$-$\bar D4$ systems with $D2$ charges
\footnote{
We should mention that the form of the conditions $(4.13), (4.14)$ in \cite{Lugo:2002wr} (or $(2.9)$ in
\cite{Lugo:2004qt}) led to one of us to assign (erroneously) Taub-Nut charges to the solution;
however it can be easily showed that they are equivalent to,
\be
\Gamma_{012}\,\epsilon = \pm\,sign (B_2)\,\epsilon\qquad;\qquad \Gamma_{034}\,\epsilon = \pm\,sign (B_1)\,\epsilon
\ee
that manifestly show up just $D2$ (or $\bar D2$) charges on both planes $(12)$ and $(34)$.
}
and no $D4$-brane charge (which would represent genuine bound states of $D2$ and $\bar D2$ branes) was conjectured.
While it is plausible that higher dimensional supertube-like solutions exist, leading in a
certain limit to brane-antibrane systems much as it happens with the supertube,
in this letter we will focus on the analysis both from the low energy point of view and in the weak coupling limit,
along the lines of references \cite{Lugo:2002wr}, \cite{Lugo:2004qt} (see also \cite{Chen:1999bf}, \cite{Chen:2000ks},
\cite{Myers:2002bk}, \cite{Bak:2001tt}), of flat $D6 -\bar D 6$ systems with constant magnetic fields on their world-volume.
We will determine the conditions under which some SUSY charges are preserved,  tachyonic instabilities are absent,
and no force between the branes is present.
\bigskip

\section{The DBI low energy action and supersymmetry.}
\cleqn

Let us consider a $D p$ -brane parameterized by world-volume coordinates $\{\xi^\mu, \mu=0,1,\dots,p\}$, in a
configuration defined by the embedding fields $\{ X^M(\xi), M=0,1,\dots,9 \}\,$ in the ten-dimensional space-time,
and the abelian gauge field $A= d\xi^\mu\,A_\mu(\xi)\;$,  $F=dA$ being the field strength.
They are the fields associated to the bosonic massless states of a $Dp$ brane in type II string theories.
The effective action for them is the DBI action
\footnote{
The pull-back's of a tensor field $T_{M_1\dots M_n}$ and gamma-matrices $\Gamma^A$ to the brane are defined by,
\bea
t_{\mu_1\dots\mu_n}(\xi) &\equiv& T_{M_1\dots M_n}(X)|_{X(\xi)}\;
\partial_{\mu_1} X^{M_1}(\xi)\dots\partial_{\mu_n} X^{M_n}(\xi)\cr
d\xi^\mu\,\gamma_\mu &\equiv&  E^A(X)|_{X(\xi)}\,\Gamma_A
\eea
where $E^A = dX^M\, E_M{}^{A}(X)$ is the ten dimensional vielbein.
Furthermore,  $\Gamma^{A_1...A_n}$ ($\gamma^{\mu_1...\mu_n}$) is the weight one antisymmetrized product of
$\Gamma^{A_1},\dots,\Gamma^{A_n}\;$ $(\gamma^{\mu_1},\dots,\gamma^{\mu_n}\;)$,
$\;\{\Gamma_A ;\Gamma_B\} = 2\,\eta_{AB}\;$
($\;\{\gamma_\mu ;\gamma_\nu\} = 2\,g_{\mu\nu}$)
, and $\Gamma_{11} = \Gamma_0\, \Gamma_1\dots\Gamma_9$.
The induced volume form on the brane is
$ \epsilon_{\mu_1\dots\mu_{p+1}} \equiv \sqrt{|g|}\, \varepsilon_{\mu_1\dots\mu_{p+1}}$,
where $\varepsilon_{01\dots p}= -1$ in some patch defines an orientation; condition (\ref{susycond}) with a $``-"$ sign on the r.h.s.
corresponds to the anti-brane with the {\it same} fields of the brane, since by definition they have opposite orientations.
},
\bea
S_{DBI}=-T_p\int d^{p+1}\xi\; e^{-\Phi}\;\sqrt{-\det(g_{\mu\nu}+b_{\mu\nu} + T_s{}^{-1}\, F_{\mu\nu})}^,\label{dbi}
\eea
where $T_s= (2\pi\alpha')^{-1}$ (put to $1$ in most places along the paper) is the string tension, and $T_p$ is the $Dp$-brane tension.
Let $\epsilon$ be the general (Majorana) Killing spinor of some background
$(G_{MN}, B_{MN}, \phi, A^{(p+1)}_{M_1\dots M_{p+1}})\,$ of type $IIA$ string theory.
Then, from the analysis of the SUSY extension of (\ref{dbi}), the introduction of the $Dp$ brane in such space will preserve
the supersymmetries that satisfy \cite{Bergshoeff:1997kr},
\be
\Gamma\;\epsilon = +\;\epsilon\label{susycond}
\ee
where the $\Gamma$-matrix is defined by \cite{Bergshoeff:1996tu},
\bea
\Gamma &\equiv& \frac{1}{ |d|^\frac{1}{2}  }\;
\sum_{n=0}^{\left[\frac{p+1}{2}\right]}\; \frac{1}{2^n\,n!}\; f_{\mu_1\nu_1}\dots
f_{\mu_n\nu_n}\;\gamma^{\mu_1\nu_1\dots\mu_n\nu_n}\; (\Gamma_{11})^{n + \frac{p-2}{2}}\;\Gamma_{(0)}\cr
\Gamma_{(0)} &\equiv&
\frac{1}{(p+1)!}\;\epsilon_{\mu_1\dots\mu_{p+1}}\;\gamma^{\mu_1\dots\mu_{p+1}}
\label{gama}
\eea
and $d\equiv \det(\delta^\mu{}_\nu + {\it f}^\mu{}_\nu)\,$, $f^\mu{}_\nu\equiv g^{\mu\rho}\, (F_{\rho\nu} + b_{\rho\nu}) =
g^{\mu\rho}\, f_{\rho\nu}$.

We are ready to start with our analysis.
We will restrict to work on the flat ten-dimensional Minkowski vacuum of type IIA string theory,
the spinor $\epsilon$ in (\ref{susycond}) being a $32$ dimensional constant one (in the standard vielbein).

\section{The $D6 -\bar D6$ system}
\cleqn

Let us consider a flat $D6$-brane extended along $(X^0,\dots ,X^6)$, with constant field strength
\footnote{
Given a constant field strength $F_{i0} = E_i\;,\; F_{ij}= B_{ij}$, we can eliminate the electric field
by means of a boost with velocity $\vec\beta = - B^{-1}\,\vec E$, provided that we restrict the fields to the region
$\beta^2= {\vec E}^t\, (-B^2)^{-1}\,\vec E<1$; a further $SO(6)$ rotation puts the field strength in the form (\ref{fd4}).
},
\be
F_{12} \equiv B_1 \qquad,\qquad F_{34} \equiv B_2\qquad,\qquad F_{56} \equiv B_3\label{fd4}
\ee
The $\Gamma$-matrix is,
\bea
\Gamma &=& \frac{1}{\sqrt{|d|}}\,\left(1 + \frac{1}{2}F_{\mu\nu}\,\Gamma^{\mu\nu}\,\Gamma_{11}
+\frac{1}{8}F_{\mu\nu}\,F_{\rho\sigma}\,\Gamma^{\mu\nu\rho\sigma}
+\frac{1}{48}F_{\mu\nu}\,F_{\rho\sigma}\,F_{\eta\lambda}\,\Gamma^{\mu\nu\rho\sigma\eta\lambda}\,\Gamma_{11}\right)\Gamma_{(0)}\cr
& &\label{gamapreliminar}
\eea
that for our gauge configuration (\ref{fd4}) can be written as,
\bea
\Gamma &=& d^{-\frac{1}{2}}\;\left(\Gamma_{0123456}- B_1\,\Gamma_{12789}-B_2\,\Gamma_{34789}-B_3\,\Gamma_{56789}+ B_1\;B_2\;\Gamma_{056} \right.\cr
&+& \left.  B_1\;B_3\;\Gamma_{034}+B_2\;B_3\;\Gamma_{012}- B_1\;B_2\;B_3\;\Gamma_{11}\;\Gamma_{0} \right)\cr
d &=& \left(1+B_1^2\right)\left(1+B_2^2\right)\left(1+B_3^2\right)
\label{gamma6}
\eea
In the following subsections we will present two solutions for systems of parallel $D6-\bar D6$ with determined configurations of fields that
will preserve four supercharges.
The method used, developed in \cite{Bak:2001xx}, \cite{Lugo:2002wr}, \cite{Lugo:2004qt}, consists in dividing (\ref{susycond})
in two or more constraints compatibles by themselves and among them.

\subsection{Solution I}

Let us consider the following operators,  $\Gamma_{\oa}=\Gamma_{0\,1\,2}$, $\Gamma_{\ob}=\Gamma_{0\,3\,4}$ and $\Gamma_{\oc}=\Gamma_{0\,5\,6}$,
and impose the following constraints,
\be
\Gamma_{\oa}\;\epsilon=\mu_1\,\epsilon\qquad,\qquad\Gamma_{\ob}\;\epsilon=\mu_2\,\epsilon\qquad,\qquad
\Gamma_{\oc}\;\epsilon=\mu_3\,\epsilon\label{ansatz1}
\ee
with $\mu_i^2=1$.
They trivially satisfy,
$\left[\Gamma_{\oa};\Gamma_{\ob}\right]=\left[\Gamma_{\oa};\Gamma_{\oc}\right]=\left[\Gamma_{\ob};\Gamma_{\oc}\right]=0\;$.
These compatibility conditions together with the property $tr\; \Gamma^{\mu_1\dots\mu_n} =0 $, guarantee the existence of
a $\frac{1}{8}$ supersymmetric solution to (\ref{susycond}).
In fact, by using (\ref{ansatz1}) in (\ref{susycond}) we get ($\widehat{B}_i\equiv\mu_i\,B_i$),
\be
\frac{1}{\sqrt{d}}\,\left(\mu_1\,\mu_2\,\mu_3\,\left(\widehat{B}_1\widehat{B}_2+\widehat{B}_2\widehat{B}_3+\widehat{B}_3\widehat{B}_1 -1\right)
+\left(\widehat{B}_1 +\widehat{B}_2 +\widehat{B}_3 -\widehat{B}_1\widehat{B}_2\widehat{B}_3\right)\,\Gamma_{0\,7\,8\,9}\right)\;\epsilon =\pm\epsilon
\label{cond1}
\ee
where $+(-)$ stands for the $D6$ ($\bar D6$) brane.
It is convenient at this point to introduce the indices $\nu^{(i)}$ (and $\widehat{\nu}^{(i)} \equiv \mu_i\,\nu^{(i)}$) through,
\be
B_i\equiv \tan\left(\frac{\pi}{2}\nu^{(i)}\right)\label{nu}\qquad,\qquad |\nu^{(i)}|<1.
\ee
Then, in view of the anticommutation of $\Gamma_{0\,7\,8\,9}$ with the $\Gamma_{\oi}$ 's, (\ref{cond1}) yields the following conditions
for the field configuration,
\be
\sin\left(\frac{\pi}{2}\left(\widehat{\nu}^{(1)} + \widehat{\nu}^{(2)} + \widehat{\nu}^{(3)} \right)\right)=0\qquad and\qquad
\cos\left(\frac{\pi}{2}\left(\widehat{\nu}^{(1)} + \widehat{\nu}^{(2)} + \widehat{\nu}^{(3)} \right)\right)=\mp\mu_1\,\mu_2\,\mu_3
\ee
with possible solutions,

\noindent{\bf $D6$ brane}
\bea
\begin{array}{l}\mu_1\mu_2\mu_3 = - 1 \cr \widehat{\nu}_b^{(1)} + \widehat{\nu}_b^{(2)} + \widehat{\nu}_b^{(3)} = 0
\end{array}
\qquad\qquad or\qquad\qquad
\begin{array}{l}
\mu_1\mu_2\mu_3 = + 1 \cr \widehat{\nu}_b^{(1)} + \widehat{\nu}_b^{(2)} + \widehat{\nu}_b^{(3)} = 2, -2
\end{array}\label{d6sn1}
\eea
\noindent{\bf $\bar D 6$ brane}
\bea
\begin{array}{l}
\mu_1\mu_2\mu_3 = - 1 \cr \widehat{\nu}_a^{(1)} + \widehat{\nu}_a^{(2)} + \widehat{\nu}_a^{(3)} = 2, -2\end{array}
\qquad or\qquad\qquad
\begin{array}{l}\mu_1\mu_2\mu_3 = + 1 \cr \widehat{\nu}_a^{(1)} + \widehat{\nu}_a^{(2)} + \widehat{\nu}_a^{(3)} = 0\end{array}
\label{bard6sn1}
\eea
If we are interested in a $D6-\bar D6$ supersymmetric system, (and stable as consequence of the BPS character) we must seek for configurations
of fields in the brane and in the antibrane that preserve the same supersymmetries, or equivalently that have the same Killing spinors.
As from (\ref{ansatz1}) this happens iff the $\mu_i{}$' s are the same in both the brane and the antibrane,
we conclude from (\ref{d6sn1}), (\ref{bard6sn1}), that a SUSY system must lie in one of the following  cases,
\bea
\mu_1\mu_2\mu_3&=&-1\qquad\longrightarrow\qquad
\widehat{\nu}_b^{(1)} + \widehat{\nu}_b^{(2)} + \widehat{\nu}_b^{(3)}=0\qquad\; and\qquad
\widehat{\nu}_a^{(1)} + \widehat{\nu}_a^{(2)} + \widehat{\nu}_a^{(3)}=2,-2\cr
\mu_1\mu_2\mu_3&=&+1\qquad\longrightarrow\qquad
\widehat{\nu}_b^{(1)} + \widehat{\nu}_b^{(2)} + \widehat{\nu}_b^{(3)}=2,-2\;\;and\qquad
\widehat{\nu}_a^{(1)} + \widehat{\nu}_a^{(2)} + \widehat{\nu}_a^{(3)}=0\cr
& &\label{d6bard6sn1}
\eea
We can explicitly get the Killing spinors in a Weyl basis $\{ (s_1,\dots,s_5)\;,\; s_i = 0,1\;\}$
of the spinorial representation (see \cite{polcho2}), where the operators $\Gamma_{\oi}$ are represented by,
\be
\Gamma_{\oa}=\left(1\,\sigma_3\,\sigma_3\,\sigma_3\,\sigma_2\right)\qquad,\qquad
\Gamma_{\ob}=\left(\sigma_3\,1\,\sigma_3\,\sigma_3\,\sigma_2\right)\qquad,\qquad
\Gamma_{\oc}=\left(\sigma_3\,\sigma_3\,1\,\sigma_3\,\sigma_2\right)
\ee
A basis where $\Gamma_{\oa}, \Gamma_{\ob}, \Gamma_{\oc}$, are diagonal is,
$\xi^{\pm}_{(s_1\,s_2\,s_3\,s_4)}=\left(s_1\,s_2\,s_3\,s_4\,1\right)\pm i \left(s_1\,s_2\,s_3\,s_4\,0\right)$,
with eigenvalues $\pm (-)^{\bar s_2 + \bar s_3 + \bar s_4}, \pm (-)^{\bar s_1 + \bar s_3 + \bar s_4}$ and
$\pm(-)^{\bar s_1 + \bar s_2 + \bar s_4}$, respectively.
According to the choice of the $\mu_i{}'s$, we have $\frac{1}{8}\,32=4$  Killing spinors for our system.
As an example, if $\mu_1=\mu_2=\mu_3=1$, and the fields are, $\nu_b^{(1)}=\nu_b^{(2)}=\nu_b^{(3)}=\frac{2}{3}$ in the brane, and
$\nu_a^{(1)}=-\nu_a^{(2)}=\frac{1}{2}, \nu_a^{(3)}=0$ in the antibrane (that
correspond to $B_1=B_2=B_3=\sqrt{3}$ in the $D6$-brana, and $B_1=-B_2=1,B_3=0$ in the $\bar D6$-brane),
the general Killing spinor is,
\bea
\epsilon=\xi_1\;\xi^{+}_{(0\,0\,0\,1)} + \xi_2\;\xi^{+}_{(1\,1\,1\,1)} + \xi_3\;\xi^{-}_{(0\,0\,0\,0)} + \xi_4\;
\xi^{-}_{(1\,1\,1\,0)}
\eea
where the $\xi_i{}$'s are complex constants subject to the Majorana condition for $\epsilon$.

\subsection{Solution II}

Now we present another solution compatible with the existence of $\frac{1}{8}$ SUSY $D6$-$\bar D 6$  systems.
Let us consider the following conditions,
\bea
\Gamma_{\oa}\epsilon &=& \epsilon\qquad,\qquad \Gamma_{\oa} \equiv -B_1\,B_2 \;\Gamma_{1234}-B_1\,B_3 \;\Gamma_{1256}-B_2\,B_3 \;\Gamma_{3456}\cr
\Gamma_{\ob}\epsilon &=& \epsilon\qquad,\qquad\Gamma_{\ob} \equiv \Gamma_{11}\;\Gamma_0\cr
\Gamma_{\oc}\epsilon &=& \pm\epsilon\;\;\;\;,\qquad
\Gamma_{\oc}\equiv-\frac{1}{\sqrt{d}}\left(B_1B_2B_3 + B_1 \Gamma_{3456} + B_2 \Gamma_{1256} + B_3 \Gamma_{1234} \right)
\label{cond2}
\eea
It is readily shown that if they hold, (\ref{susycond}) follows; also the consistency condition
\footnote{
The relations, $[\Gamma_{\oa};\Gamma_{\ob}]= [\Gamma_{\oa};\Gamma_{\oc}]=[\Gamma_{\ob};\Gamma_{\oc}]=0$
and $\Gamma_{\ob}{}^2\epsilon = \epsilon$ are straightforward.
},
\be
\Gamma_{\oc}^2\,\epsilon = \left(1 - d^{-1}\left(1-\Gamma_{\oa}\right)^2\right)\,\epsilon = \epsilon\label{gammac2}
\ee
holds. On the other hand,
\be
\Gamma_{\oa}^2\,\epsilon=\left(1 + \frac{-1 + \sum_{i=1}^{3} \sin^2(\frac{\pi}{2}\nu^{(i)}) \pm 2\,
\prod_{i=1}^{3}\sin(\frac{\pi}{2}\nu^{(i)})}{\cos^2(\frac{\pi}{2}\nu^{(1)}) \cos^2(\frac{\pi}{2}\nu^{(2)}) \cos^2(\frac{\pi}{2}\nu^{(3)})}\right)\,
\epsilon=\epsilon\label{gammaa2}
\ee
yields the following constraint on the fields,
\be
\sin^2\left(\frac{\pi}{2}\nu^{(1)}\right) + \sin^2\left(\frac{\pi}{2}\nu^{(2)}\right)+ \sin^2\left(\frac{\pi}{2}\nu^{(3)}\right)
\pm 2\, \sin(\frac{\pi}{2}\nu^{(1)})\,\sin(\frac{\pi}{2}\nu^{(2)})\,\sin(\frac{\pi}{2}\nu^{(3)}) -1=0
\ee
which is solved by,

\noindent{\bf $D6$ brane}
\be
\sin\left(\frac{\pi}{2}\nu_b^{(3)}\right) = \pm\, \cos\left(\frac{\pi}{2}(\nu_b^{(1)}\pm\nu_b^{(2)}\right)\qquad\longrightarrow\qquad
\begin{array}{l}
\mp\nu^{(1)}_b\mp\nu^{(2)}_b+\nu^{(3)}_b =1\qquad or\cr
\pm\nu^{(1)}_b\mp\nu^{(2)}_b-\nu^{(3)}_b=1
\end{array}\label{d6sn2}
\ee
\noindent{\bf $\bar D 6$ brane}
\be
\sin\left(\frac{\pi}{2}\nu_a^{(3)}\right) = \pm\, \cos\left(\frac{\pi}{2}(\nu_a^{(1)}\mp\nu_a^{(2)}\right)\qquad\longrightarrow\qquad
\begin{array}{l}
\mp\nu_a^{(1)}\pm\nu_a^{(2)}+\nu_a^{(3)} =1\qquad or\cr
\pm\nu_a^{(1)}\pm\nu_a^{(2)}-\nu_a^{(3)}=1
\end{array}\label{bard6sn2}
\ee
As made with the Solution I, we can explicitly get the Killing spinors, solutions of (\ref{cond2}).
Let $\{\epsilon^{\pm}_{(s_1s_2s_3s_4)}= \left(s_1\,s_2\,s_3\,s_4\,0\right) \pm \left(s_1\,s_2\,s_3\,s_4\,1\right)\}\,$ be
a basis in which $\Gamma_{\ob}$ is diagonal with eigenvalues $\pm 1$.
Then, as an example, both a brane with $\sum_{i=1}^3\,\nu^{(i)}_b = +1$ and an antibrane with $\sum_{i=1}^3\,\nu^{(i)}_a=-1$,
have the general Killing spinor,
\be
\epsilon = \varsigma_1\,\epsilon^+_{(0\,0\,0\,0)} + \varsigma_2\,\epsilon^+_{(0\,0\,0\,1)} + \varsigma_3\,\epsilon^+_{(1\,1\,1\,0)} +
\varsigma_4\,\epsilon^+_{(1\,1\,1\,1)}\label{killingsn2}
\ee
Similarly we can solve (\ref{cond2}) in the other cases of (\ref{d6sn2}), (\ref{bard6sn2}) obtaining that, in order
to have $D6-\bar D6$ SUSY systems, i.e. the same Killing spinors for branes that antibranes, one of the four relations must hold,
\bea
+\nu_b^{(1)}+\nu_b^{(2)}+\nu_b^{(3)}=1\qquad&\emph{and}&\qquad -\nu_a^{(1)}-\nu_a^{(2)}-\nu_a^{(3)}=1\qquad or\cr
-\nu_b^{(1)}-\nu_b^{(2)}+\nu_b^{(3)}=1\qquad&\emph{and}&\qquad +\nu_a^{(1)}+\nu_a^{(2)}-\nu_a^{(3)}=1\qquad or\cr
-\nu_b^{(1)}+\nu_b^{(2)}-\nu_b^{(3)}=1\qquad&\emph{and}&\qquad +\nu_a^{(1)}-\nu_a^{(2)}+\nu_a^{(3)}=1\qquad or\cr
+\nu_b^{(1)}-\nu_b^{(2)}-\nu_b^{(3)}=1\qquad&\emph{and}&\qquad -\nu_a^{(1)}+\nu_a^{(2)}+\nu_a^{(3)}=1\label{d6bard6sn2}
\eea
the first one corresponding to (\ref{killingsn2}).

\section{Weak coupling analysis}
\cleqn

In this section, we will analyze the perturbative spectrum of the $D6$-$\bar D6$ system
defined in Section $3$, as well as the one loop amplitude.
We will focus on the inter-brane sector, so we must consider open superstrings suspended between the
$D6$-brane and the $\bar D6$-brane, with a time-like NN coordinate $X^0$,
six coordinates along the branes resumed in three complex fields
$Z^{(1)} \equiv X^1 + i X^2, Z^{(2)}\equiv X^3 + iX^4 , Z^{(3)}\equiv X^5 + iX^6$, obeying the b.c.,
\be
\partial_\sigma X^\mu(\tau,0) - f_0{}^\mu{}_\nu\;\partial_\tau X^\nu(\tau,0)
=\partial_\sigma X^\mu(\tau,\pi) - f_\pi{}^\mu{}_\nu\;\partial_\tau
X^\nu(\tau,\pi)= 0
\;\;\;.\label{oxbc}
\ee
coming from the coupling of the ends of the strings (at $\sigma=0$ and $\sigma=\pi$) to the gauge field $F_0{}^\mu{}_\nu = T_s\,f_0{}^\mu{}_\nu$
given in (\ref{fd4}), and three DD coordinates $X^i , i=7, 8, 9$ orthogonal to the branes.
Each coordinate field is paired with fermionic partners $\psi^0$ (Majorana),
$\Psi^{(1)} ,\Psi^{(2)}, \Psi^{(3)}$ (Dirac's) and $\psi^i,i=7, 8, 9\,$ (Majoranas) respectively, with equal (one-half shifted) modding as
its bosonic partner in the R (NS) sector
\footnote{
This section does not pretend to be self-contained; for the expansions, notation, etc., we heavily refer the reader to the appendix
of \cite{Lugo:2004qt}.
}.

\subsection{Analysis of the spectrum and supersymmetry}

For simplicity, we will work in the light-cone gauge.
In reference \cite{Lugo:2004qt} was showed that, although $X^0$ and $X^9$ have different boundary conditions (NN and DD respectively), it is consistent
to work with light-cone coordinates $X^{\pm}=X^0\pm X^{9}$, that obey the following b.c.,
\be
\partial_{\pm}X^+|_{\sigma=0}=\partial_{\mp}X^-|_{\sigma=0}\qquad,\qquad
\partial_{\pm}X^+|_{\sigma=\pi}=\partial_{\mp}X^-|_{\sigma=\pi}\;,\label{lcbc}
\ee
The mode expansions for them results,
\be
X^{\pm}=x^{\pm}+ \alpha'\;p^\pm\;\sigma^+ + i\,\frac{l}{2}\;\sum_{m\in\bz'}\; \frac{\alpha_m^\pm}{m} \; e^{-im\sigma^+} +
\alpha'\;p^\mp\;\sigma^- + i\,\frac{l}{2}\;\sum_{m\in\bz'}\;\frac{\alpha_m^\mp}{m} \; e^{-im\sigma^-}
\ee
where $p^{\pm}\equiv E\pm T_s\Delta x^9$, $E\equiv p^0$, and $\Delta x^9$ is the separation between branes along the direction $9$.
On the other hand, the companion fermions are
$\psi_R^\pm \equiv \psi_R^0\pm\psi_R^9 \sim\sum_{r\in \bz_\nu}\;b^\pm_r\; e^{-ir\sigma^\pm}$,
with $\nu=0\, (\frac{1}{2})$ in R (NS) sectors, and similarly for the left fermions.
The light cone gauge is carried out by fixing $\{\alpha_m^+=b^+_m =0\;,\;m\neq 0\}$, and the spectra is constructed from the transverse
oscillators and zero modes.
The mass shell condition results,
\be
\alpha'\,E^2 = \frac{L^2}{4\pi^2\alpha'} + N^\bot + \delta\left( \bar\nu^{(1)} + \bar\nu^{(2)} + \bar\nu^{(3)} - 1 \right)
\ee
where $L\equiv\sqrt{(\Delta x^7){}^2+(\Delta x^8){}^2+(\Delta x^9){}^2}$ is the separation between branes (put eventually to zero),
$N^\bot$ is the total transverse oscillator energy operator and,
\be
\nu^{(i)}_- \equiv \pm\frac{1}{2}\,(\nu^{(i)}_b - \nu^{(i)}_a)\qquad \bar d d /d\bar d \qquad,\qquad
\bar\nu^{(i)} = \left\{\begin{array}{lcr}\nu^{(i)}_-\;\;\;&,&\;\;\; 0\leq\nu^{(i)}_-<1\cr\nu^{(i)}_- +1\;\;\;&,&\;\;\;
-1<\nu^{(i)}_-<0\end{array}\right.\;\;\;\;.\label{barnu}
\ee
By taking into account the GSO projection, that preserves only the states that satisfy,
\bea
(-)^{ \sharp + \nu_-^{(1)} - \bar\nu^{(1)} + \nu_-^{(2)} -
\bar\nu^{(2)}+ \nu_-^{(3)} - \bar\nu^{(3)}}\left\{\begin{array}{lcr}1 \cr \gamma
\end{array}\right.\;\;\;\;= +1\qquad
\begin{array}{l}NS \cr R \end{array}\label{GSObruto}
\eea
where $\sharp$ is the (world sheet) fermion number and $\gamma\equiv -2\,i\, b_0^7\,b_0^8\,\,$ is the chirality operator of the
$spin(2)$ transverse group (the operators $P_+ (P_-)$ introduced below are the chirality positive (negative) projectors)
\footnote{
We remember that due to the mixed coordinates there is no zero mode $b_0^j$, j$=1,\dots,6$,
and therefore there is no contribution to the vacuum degeneration.
},
we have found the following spectra for the lowest levels.
\bigskip

\noindent{\bf Spectrum solution I}

For definiteness, we fix $\mu_1=\mu_2=\mu_3=1$, and $\nu_b^{(1)}+\nu_b^{(2)}+\nu_b^{(3)}=2$
in the brane and $\nu_a^{(1)}+\nu_a^{(2)}+\nu_a^{(3)}=0$ in the antibrane
(see (\ref{d6bard6sn1})).
The $\nu_b^{i}$'s , $i=1,2,3$, are necessarily positive, and we furthermore take
$0<\nu_a^{(1)}<\nu_b^{(1)},\,\nu_a^{(2)}<0,\, \nu_a^{(3)}>\nu_b^{(3)}$, the other cases having similar spectra.
At left (right) are written the space-time bosons (fermions), and $|\alpha>\,,\,\alpha=1,2\;$, stands for the $spin(2)$ spinor state.
\bigskip
\begin{itemize}

\item\underline{Fundamental level}$\qquad\alpha' \, E^2 = \frac{T_s}{2\pi}\, L^2$

Sector $d\bar d$
\be
|0>_{NS}\qquad,\qquad  P_+|\alpha>
\ee

Sector $\bar d d$
\be
B^{(1)}_{\bar\nu^{(1)}-\frac{1}{2}}\;B^{(2)}_{\bar\nu^{(2)}-\frac{1}{2}}\; B^{(3)}_{\bar\nu^{(3)}-\frac{1}{2}} |0>_{NS}\qquad,\qquad
P_-|\alpha>
\ee
We see that the lowest level has two bosonic and two fermionic degrees of freedom, that transforms as scalars and a spin one-half representations
of the unbroken group $\overline{SO(3)}\sim SU(2)$.
Let us go to the first excited levels.
\bigskip

\item\underline{Level}   $\alpha'\,E^2=  \frac{T_s}{2\pi}\, L^2 + |\nu_-^{(1)}|$

Sector $d\bar d$
\bea
\begin{array}{r}
A^{(1)}_{\bar\nu^{(1)}-1}|0>_{NS}\cr
A^{(3)}_{\bar\nu^{(3)}}{}^\dagger\;A^{(2)}_{\bar\nu^{(2)}}{}^\dagger|0>_{NS}\cr
b^L_{-\frac{1}{2}}\;B^{(1)}_{\bar\nu^{(1)}-\frac{1}{2}}|0>_{NS},\;\;L=7,8.\cr
A^{(L)}_{\bar\nu^{(L)}}{}^\dagger\;B^{(L)}_{\bar\nu^{(L)}-\frac{1}{2}}\;B^{(1)}_{\bar\nu^{(1)}-\frac{1}{2}}|0>_{NS},\;\;L=2,3
\end{array}\qquad;\qquad
\begin{array}{rl}
A^{(1)}_{\bar\nu^{(1)}-1}P_+|\alpha>,\cr
A^{(3)}_{\bar\nu^{(3)}}{}^\dagger\;A^{(2)}_{\bar\nu^{(2)}}{}^\dagger P_+|\alpha>,\cr
B^{(1)}_{\bar\nu^{(1)}-1} P_-|\alpha>,\cr
A^{(3)}_{\bar\nu^{(3)}}{}^\dagger\;B^{(2)}_{\bar\nu^{(2)}}{}^\dagger P_-|\alpha>,\cr
A^{(2)}_{\bar\nu^{(2)}}{}^\dagger\;B^{(3)}_{\bar\nu^{(3)}}{}^\dagger P_-|\alpha>,\cr
B^{(3)}_{\bar\nu^{(3)}}{}^\dagger\;B^{(2)}_{\bar\nu^{(2)}}{}^\dagger P_+|\alpha>,
\end{array}
\eea

Sector $\bar dd $
\bea
\begin{array}{r}
A^{(3)}_{\bar\nu^{(3)}-1}\;B^{(2)}_{\bar\nu^{(2)}-\frac{1}{2}}|0>_{NS}\cr
A^{(2)}_{\bar\nu^{(2)}-1}\;B^{(3)}_{\bar\nu^{(3)}-\frac{1}{2}}|0>_{NS}\cr
A^{(1)}_{\bar\nu^{(1)}}{}^\dagger\;B^{(1)}_{\bar\nu^{(1)}-\frac{1}{2}}\;B^{(2)}_{\bar\nu^{(2)}-\frac{1}{2}}\;B^{(3)}_{\bar\nu^{(3)}-
\frac{1}{2}}|0>_{NS}\cr
A^{(3)}_{\bar\nu^{(3)}-1}\;A^{(2)}_{\bar\nu^{(2)}-1}\;B^{(1)}_{\bar\nu^{(1)}-\frac{1}{2}}\;B^{(2)}_{\bar\nu^{(2)}-\frac{1}{2}}\;B^{(3)}_{\bar\nu^{(3)}
-\frac{1}{2}}|0>_{NS}\cr
b^L_{-\frac{1}{2}}\;B^{(3)}_{\bar\nu^{(3)}-\frac{1}{2}}\;B^{(2)}_{\bar\nu^{(2)}-\frac{1}{2}}|0>_{NS},\;\;L=7,8.\cr
\end{array}\qquad;\qquad
\begin{array}{rl}
A^{(1)}_{\bar\nu^{(1)}}{}^\dagger P_-|\alpha>,\cr
A^{(3)}_{\bar\nu^{(3)}-1}\;A^{(2)}_{\bar\nu^{(2)}-1} P_-|\alpha>,\cr
B^{(1)}_{\bar\nu^{(1)}}{}^\dagger P_+|\alpha>,\cr
A^{(3)}_{\bar\nu^{(3)}-1}\;B^{(2)}_{\bar\nu^{(2)}-1} P_+|\alpha>,\cr
A^{(2)}_{\bar\nu^{(2)}-1}\;B^{(3)}_{\bar\nu^{(3)}-1} P_+|\alpha>,\cr
B^{(3)}_{\bar\nu^{(3)}-1}\;B^{(2)}_{\bar\nu^{(2)}-1} P_-|\alpha>,
\end{array}
\eea
This level presents 12 bosons and 12 fermions.
\bigskip

\item\underline{Level}   $\alpha'\,E^2=  \frac{T_s}{2\pi}\, L^2 + |\nu_-^{(2)}|$

The states coincide with those of the former case by exchanging $\bar\nu^{(1)}\leftrightarrow \bar\nu^{(2)}$.
\bigskip

\item\underline{Level}   $\alpha'\,E^2=  \frac{T_s}{2\pi}\, L^2 + |\nu_-^{(3)}|$

Sector $d\bar d$
\bea
\begin{array}{r}
A^{(3)}_{\bar\nu^{(3)}}{}^\dagger|0>_{NS},\cr
B^{(1)}_{\bar\nu^{(1)}-\frac{1}{2}}\;B^{(2)}_{\bar\nu^{(2)}-\frac{1}{2}}|0>_{NS},
\end{array}\qquad;\qquad
\begin{array}{rl}
A^{(3)}_{\bar\nu^{(3)}}{}^\dagger P_+|\alpha>,\cr
B^{(3)}_{\bar\nu^{(3)}}{}^\dagger P_-|\alpha>,
\end{array}
\eea

Sector $\bar dd$
\bea
\begin{array}{r}
B^{(3)}_{\bar\nu^{(3)}-\frac{1}{2}}|0>_{NS},\cr
A^{(3)}_{\bar\nu^{(3)}-1}\;B^{(1)}_{\bar\nu^{(1)}-\frac{1}{2}}\;B^{(2)}_{\bar\nu^{(2)}-\frac{1}{2}}\;B^{(3)}_{\bar\nu^{(3)} -\frac{1}{2}}|0>_{NS},
\end{array}\qquad;\qquad
\begin{array}{rl}
A^{(3)}_{\bar\nu^{(3)}-1} P_-|\alpha>,\cr
B^{(3)}_{\bar\nu^{(3)}-1} P_+|\alpha>,
\end{array}
\eea
\end{itemize}
This level presents $4$ bosons and $4$ fermions.
\bigskip

\noindent{\bf Spectrum solution II}

We present here the spectrum for the lowest levels in the first case of (\ref{d6bard6sn2}),
\be
\nu_-^{(1)}+\nu_-^{(2)}+\nu_-^{(3)}=\pm1\qquad,\qquad \bar d d/ d \bar d
\ee
We distinguish two different subcases, depending on whether all $\nu_-^{(i)}$'s are positive (negative), or one of them is negative (positive) and the
other two are positive (negative) in the $\bar dd$ ($d\bar d$) sector ($\nu_-^{(i)}\neq 0,i=1,2,3$, is assumed).
The other cases are similar in field content.
\bigskip

\noindent{\bf Case 1} $\;\nu_a^{(i)}<\nu_b^{(i)}\;,\; i=1,2,3\;$
\bigskip

\begin{itemize}

\item\underline{Fundamental level}$\qquad\alpha' \, E^2 = \frac{T_s}{2\pi}\, L^2$

Sector $\bar dd$
\be
|0>_{NS}\qquad,\qquad P_+|\alpha>
\ee
Sector $ d\bar d$
\be
B^{(1)}_{\bar\nu^{(1)}-\frac{1}{2}}\;B^{(2)}_{\bar\nu^{(2)}-\frac{1}{2}}\; B^{(3)}_{\bar\nu^{(3)}-\frac{1}{2}}|0>_{NS}
\qquad,\qquad P_-|\alpha>,
\ee

We see that the lowest level has two bosonic and two fermionic degrees of freedom, that transforms respectively
as scalars and a spin one-half representations of $spin(3)$.
\bigskip

\item\underline{Level}   $\alpha'\,E^2=  \frac{T_s}{2\pi}\, L^2 + |\nu_-^{(1)}|$

Sector $\bar dd$
\bea
\begin{array}{r}
A^{(1)}_{\bar\nu^{(1)}}{}^\dagger|0>_{NS},\cr
B^{(2)}_{\bar\nu^{(2)}-\frac{1}{2}}\;B^{(3)}_{\bar\nu^{(3)}-\frac{1}{2}}|0>_{NS}
\end{array}\qquad;\qquad
\begin{array}{rl}
A^{(1)}_{\bar\nu^{(1)}}{}^\dagger P_+|\alpha>,\cr
B^{(1)}_{\bar\nu^{(1)}}{}^\dagger P_-|\alpha>
\end{array}
\eea

Sector $d \bar d $
\bea
\begin{array}{r}
B^{(1)}_{\bar\nu^{(1)}-\frac{1}{2}}|0>_{NS},\cr
A^{(1)}_{\bar\nu^{(1)}-1}\; B^{(1)}_{\bar\nu^{(1)}-\frac{1}{2}}\;
B^{(2)}_{\bar\nu^{(2)}-\frac{1}{2}}\; B^{(3)}_{\bar\nu^{(3)}-\frac{1}{2}}
|0>_{NS}
\end{array}\qquad;\qquad
\begin{array}{rl}
A^{(1)}_{\bar\nu^{(1)}-1} P_-|\alpha>,\cr
B^{(1)}_{\bar\nu^{(1)}-1} P_+|\alpha>,
\end{array}
\eea

This level presents 4 bosons and 4 fermions.

\item\underline{Level}   $\alpha'\,E^2=  \frac{T_s}{2\pi}\, D^2 + |\nu_-^{(i)}|,\;i=2,3.$

Idem the former case with the exchanging $\bar\nu^{(1)} \leftrightarrow\bar\nu^{(i)} $.

\end{itemize}
\bigskip

\noindent{\bf Case 2} $\;\nu_a^{(i)}<\nu_b^{(i)}\;,i=2,3\qquad;\qquad\nu_b^{(1)}<\nu_a^{(1)}$
\bigskip

\begin{itemize}

\item\underline{Fundamental level} $\; \alpha'\,E^2=\frac{T_s}{2\,\pi}\, L^2$

Idem the fundamental level of the former case with the exchanging $d\bar d \leftrightarrow\bar dd $
\bigskip

\item\underline{Level} $\;\alpha'E^2=\frac{T_s}{2\,\pi}\, L^2 + |\nu_-^{(1)}|$

Idem the former case with the exchanging $d\bar d \leftrightarrow\bar dd $
\bigskip

\item\underline{Level} $\;\alpha'E^2=\frac{T_s}{2\,\pi}\, L^2+|\nu_-^{(2)}|$
\bigskip

Sector $\bar dd$
\bea
\begin{array}{r}
A^{(1)}_{\bar\nu^{(1)}-1}\; B^{(3)}_{\bar\nu^{(3)}-\frac{1}{2}}|0>_{NS}\cr
A^{(3)}_{\bar\nu^{(3)}-1}\; B^{(1)}_{\bar\nu^{(1)}-\frac{1}{2}}|0>_{NS}\cr
b_{-\frac{1}{2}}^L\;B^{(1)}_{\bar\nu^{(1)}-\frac{1}{2}}\;
B^{(3)}_{\bar\nu^{(3)}-\frac{1}{2}}|0>_{NS},\;\;L=7,8,\cr
A^{(2)}_{\bar\nu^{(2)}-1}\; B^{(1)}_{\bar\nu^{(1)}-\frac{1}{2}}\;
B^{(2)}_{\bar\nu^{(2)}-\frac{1}{2}}\; B^{(3)}_{\bar\nu^{(3)}-\frac{1}{2}}|0>_{NS}\cr
A^{(1)}_{\bar\nu^{(1)}-1}\; A^{(3)}_{\bar\nu^{(3)}-1}\;
B^{(1)}_{\bar\nu^{(1)}-\frac{1}{2}}\; B^{(2)}_{\bar\nu^{(2)}-\frac{1}{2}}\;
B^{(3)}_{\bar\nu^{(3)}-\frac{1}{2}} |0>_{NS}
\end{array}\qquad;\qquad
\begin{array}{rl}
A^{(2)}_{\bar\nu^{(2)}}{}^\dagger P_-|\alpha>\cr
A^{(1)}_{\bar\nu^{(1)}-1}\; A^{(3)}_{\bar\nu^{(3)}-1} P_-|\alpha>\cr
B^{(2)}_{\bar\nu^{(2)}}{}^\dagger P_+|\alpha>\cr
A^{(1)}_{\bar\nu^{(1)}-1}\; B^{(3)}_{\bar\nu^{(3)}-1} P_+|\alpha>\cr
A^{(3)}_{\bar\nu^{(3)}-1}\; B^{(1)}_{\bar\nu^{(1)}-1} P_+|\alpha>\cr
B^{(1)}_{\bar\nu^{(1)}-1}\; B^{(3)}_{\bar\nu^{(3)}-1} P_-|\alpha>
\end{array}
\eea

Sector $d \bar d$
\bea
\begin{array}{r}
A^{(2)}_{\bar\nu^{(2)}-1}|0>_{NS}\cr
A^{(1)}_{\bar\nu^{(1)}}{}^\dagger\;A^{(3)}_{\bar\nu^{(3)}}{}^\dagger|0>_{NS}\cr
b_{-\frac{1}{2}}^L\; B^{(2)}_{\bar\nu^{(2)}-\frac{1}{2}}|0>_{NS},\;\;L=7,8\cr
A^{(i)}_{\bar\nu^{(i)}}{}^\dagger\;B^{(i)}_{\bar\nu^{(i)}-\frac{1}{2}}\;
B^{(2)}_{\bar\nu^{(2)}-\frac{1}{2}}\;\;,\;\; i=1,3
\end{array}\qquad;\qquad
\begin{array}{rl}
A^{(2)}_{\bar\nu^{(2)}-1} P_+|\alpha>\cr
A^{(1)}_{\bar\nu^{(1)}}{}^\dagger\; A^{(3)}_{\bar\nu^{(3)}}{}^\dagger P_+|\alpha>\cr
B^{(2)}_{\bar\nu^{(2)}-1} P_-|\alpha>\cr
A^{(1)}_{\bar\nu^{(1)}}{}^\dagger\; B^{(3)}_{\bar\nu^{(3)}}{}^\dagger P_-|\alpha>\cr
A^{(3)}_{\bar\nu^{(3)}}{}^\dagger\; B^{(1)}_{\bar\nu^{(1)}}{}^\dagger P_-|\alpha>\cr
B^{(1)}_{\bar\nu^{(1)}}{}^\dagger\; B^{(3)}_{\bar\nu^{(3)}}{}^\dagger P_+|\alpha>
\end{array}
\eea
This level presents $12$ bosons and $12$ fermions.
\bigskip

\item $\alpha'E^2=\frac{T_s}{2\,\pi}\, L^2 + |\nu_-^{(3)}|$

Idem the last level exchanging $2\leftrightarrow 3\;,\;\bar\nu^{(2)}\leftrightarrow\bar\nu^{(3)}$.
\bigskip
\end{itemize}

We see the matching in all the levels of the bosonic and fermionic degrees of freedom, necessary condition for SUSY to hold; in particular
the fundamental level, that is massless for coincident branes, has two bosons and two fermions, having an enhancement of the degeneration if
some $B_i$ in the brane coincides with that in the antibrane, or if some of them is null.

\subsection{The one loop amplitude in the open string channel}

The {\it one-loop} diagram is constructed by imposing conditions of periodicity (P) in the euclidean time.
This is carried out by taking the traces in the Hilbert space, remembering that,
in the case of fermions and ghost system with P b.c., we must insert the spinor number operators
$(-)^{F^{\Psi}}$ and $(-)^{F^{bc}}$ respectively.
The  $\lambda=2$ ghost fields $b$-$c$ and $\lambda =\frac{3}{2}$ superghost
fields $\beta$-$\gamma$ follow the b.c. of the (bosonic) reparameterization and (fermionic)
SUGRA transformations parameters respectively.
The insertion of the spinor number operator $(-)^{F^{bc}}\; ((-)^{F^{\beta\gamma}})\,$
must be carried out when P (AP) b.c. apply, due to the fermionic (bosonic)
character of the ghost (superghost) system.

The connected part of the one loop amplitude follows from the Coleman-Weinberg formula,
\be
{\it A}^{1l} \equiv \ln\;Z^{1l}\sim -\frac{1}{2}\;tr\;(-)^{\bf F}\ln\;G^{-1}
\ee
where $G^{-1}= p^2 + M^2 = \alpha'^{-1}\,L_0$ is the inverse (free) propagator, ${\bf F}$
is the \textit{space-time} fermion number and the traces are on the full Hilbert space.
Regulating as usual the logarithm, we define the one loop amplitude as follows ($\tau\equiv i\,t$),
\bea
{\cal A}^{1l} &=& -\frac{1}{2}\;tr_{NS}\; \ln\frac{G^{-1}}{T_s}
+\frac{1}{2}\;tr_{R}\; \ln\frac{G^{-1}}{T_s} =
\int_0^\infty\;\frac{dt}{2\,t}\;\left(A_{NS}^{1l}(it) +A_R^{1l}(it)\right)\cr
A_{NS,R}^{1l}(\tau)&=&  tr_{NS,R}\; q^{L_0}\;(-)^{F^{\Psi}+ F^{bc}}|_{q=e^{i2\pi\tau}}
= A^{1l}_{(b)}(\tau)\; A^{1l}_{(f)}(\tau)|_{NS,R}\label{ampop}
\eea
The bosonic contribution to the partition function is given by,
\bea
& &A^{1l}_{(b)}(it) \equiv Z^{X^0}(\tau)\; \prod_{i=7}^9\, Z^{X^i}(\tau)\;
\prod_{j=1}^3\,Z^{Z^{(j)}}(\tau)\;
Z^{bc}(\tau)\cr
&=& -\, V_7\; e^{i\pi (1-\bar\nu^{(1)}-\bar\nu^{(2)}-\bar\nu^{(3)})}\frac{16\,\prod_{j=1}^3(b_0^{(j)}-b_\pi^{(j)})}
{(8\pi^2\alpha')^\frac{7}{2}}\;\frac{e^{-T_s\, L^2\,t\,}}{t^{\frac{1}{2}}\;\eta(it)^2}\;\prod_{j=1}^3
\left(Z^{1-2\bar\nu^{(j)}}_1 (it)\right)^{-1}\;\;,\label{ampbop}
\eea
where $V_n$ stands for the $\textbf{R}^n$-volume.
\footnote{
The zero mode of $Z^{bc}$ must be removed \cite{polcho2}.
}
In the fermionic sector we must project GSO; to do so we must insert the GSO operator when computing the trace \cite{Lugo:2004qt},
\be
P_{GSO} \equiv \frac{1}{2}\;\left( 1 -
(-)^{\nu_-^{(1)}+\nu_-^{(2)}+\nu_-^{(3)}}\;(-)^{F^{\Psi}
+F^{\beta\gamma}}\right)\;\;\;\;.\label{gsoopen}
\ee
So the fermionic contribution to the partition function in the $d\bar d$ sector results,
\bea
A^{1l}_{(f)}(it) &=& A^{1l}_{(f);GSO}(it)|_{NS} +A^{1l}_ {(f);GSO}(it)|_R\cr
A^{1l}_{(f);GSO}(it)|_{NS} &\equiv & tr_{NS}\; \prod_{a=0}^4\;
q^{L_0^{\Psi^{(a)}}-\frac{1}{24}}\;(-)^{F^{\Psi} + \sum_{a=0}^4 q_0^{(a)}}\;
q^{L_0^{\beta\gamma}-\frac{11}{24} }\; P_{GSO}\cr
&=& e^{i\pi(\bar\nu^{(1)} + \bar\nu^{(2)} +  \bar\nu^{(3)})}\frac{1}{2}
\left(Z_1^0(it)\; Z_1^{-2\bar\nu^{(1)}}(it)\;
Z_1^{-2\bar\nu^{(2)}}(it)\;Z_1^{-2\bar\nu^{(3)}}(it) \right.\cr
&+& \left. e^{-i\pi(\nu_-^{(1)} + \nu_-^{(2)} + \nu_-^{(3)})}  Z_0^0(it)\;
Z_0^{-2\bar\nu^{(1)}}(it)\; Z_0^{-2\bar\nu^{(2)}}(it)\;
Z_0^{-2\bar\nu^{(3)}}(it) \right)\cr
A^{1l}_{(f);GSO}(it)|_{R} &\equiv & tr_R\; \prod_{a=0}^4\;
q^{L_0^{\Psi^{(a)}}-\frac{1}{24}}\;(-)^{F^{\Psi} + \sum_{a=0}^4 q_0^{(a)}}\;  q^{L_0^{\beta\gamma}-
\frac{11}{24} }\;P_{GSO}\cr
&=& -\frac{1}{2}e^{i\pi(\bar\nu^{(1)} + \bar\nu^{(2)} + \bar\nu^{(3)})}
\left[Z_1^1(it)\; Z_1^{1-2\bar\nu^{(1)}}(it)\; Z_1^{1-2\bar\nu^{(2)}}(it)\;
Z_1^{1-2\bar\nu^{(3)}}(it) \right.\cr
&+& \left. e^{-i\pi(\nu_-^{(1)} + \nu_-^{(2)} + \nu_-^{(3)})}  Z_0^1(it)\;Z_0^{1-2\bar\nu^{(1)}}(it)\;
Z_0^{1-2\bar\nu^{(2)}}(it)\;Z_0^{1-2\bar\nu^{(3)}}(it)\right]\label{ampferm}
\eea
Using the periodicity property $Z_b^{a+2n}=Z_b^a\, ,\, n\in Z\;$, it is easy to see that the amplitude is invariant under
$\nu_-^{(i)}\rightarrow -\nu_-^{(i)}$,
so both the $d\bar d$ and $\bar d d$ sectors contribute the same way.

Using the \textit{second addition theorem} for theta functions \cite{fay} which states that,
\be
\prod_{i=1}^4\; \vartheta\left[\matrix{ a_i\cr b_i}\right](\nu_i;\tau)
= \frac{1}{2}\; \sum_{s_1,s_2=0,\frac{1}{2}}\;e^{-i4\pi a_1 s_2 }\;
\prod_{i=1}^4\; \vartheta\left[\matrix{ m_i + s_1\cr n_i
+s_2}\right](\epsilon_i;\tau)
\ee
holds, where,
\be
\vec\nu = J\,\vec\epsilon\qquad,\qquad \left[\matrix{\vec a \cr \vec b}\right] = J\otimes {\bf 1}_2\, \left[\matrix{\vec m \cr \vec n}\right]
\qquad,\qquad J\equiv \frac{1}{2}\;\left(\matrix{1&1&1&1\cr1&1&-1&-1\cr1&-1&1&-1\cr1&-1&-1&1\cr}\right)
\ee
with the particular choice $\epsilon_i = 0$ and the spin structures
$m_i = -\nu_-^{(i)}\;,i=1,2,3\;,\; n_1 = 1\; ,\;$ and the other zero, the fermionic amplitude (\ref{ampferm}) can be written as,
\be
A^{1l}_{(f)}(it) = \xi\;
Z_1^{-\nu_-^{(1)}-\nu_-^{(2)}-\nu_-^{(3)}}(it)\;
Z_1^{-\nu_-^{(1)}-\nu_-^{(2)}+\nu_-^{(3)}}(it)\;
Z_1^{-\nu_-^{(1)}+\nu_-^{(2)}-\nu_-^{(3)}}(it)\;
Z_1^{-\nu_-^{(1)}+\nu_-^{(2)}+ \nu_-^{(3)}}(it)\label{amplcond}
\ee
with $\xi = e^{i\pi\sum_{i=1}^3(\bar\nu^{(i)}-\nu^{(i)}_-)}$.
In virtue of the identity $Z^{1+2n}_1(it)\equiv 0\, ,\, n\in Z\;$, and taking into account
the range of definition of the variables $\nu_-^{(i)}$, (\ref{amplcond}) (and therefore the amplitude (\ref{ampop}))
is identically zero iff one of the following four constraints over the world-volume
magnetic field configuration holds,
\bea
|\nu_-^{(1)}| + |\nu_-^{(2)}| \pm |\nu_-^{(3)}| =1\qquad,\qquad
\pm |\nu_-^{(1)}| \mp|\nu_-^{(2)}| + |\nu_-^{(3)}| =1\label{nullampcond}
\eea
It is straightforward to verify that the SUSY solutions found in Section $3$, equations
(\ref{d6bard6sn1}) and (\ref{d6bard6sn2}), correspond to field configurations included in (\ref{nullampcond}).

\section{Conclusions.}
\cleqn

We have analyzed systems of $D6$ and $\bar D6$ branes with non zero magnetic fields in their world-volume both
with the DBI low energy action for the branes, and in the weak coupling, CFT limit.
In the first framework, we have found two families of conditions on the gauge fields, both of them preserving each one $\frac{1}{8}$ of the
supersymmetries of the vacuum of the type IIA superstring theory.
From the conditions (\ref{ansatz1}) we can associate to the first family $D2$ (or $\bar D2$) charges in the planes $(12), (34), (56)$,
and think of the brane-antibrane system as bound states of $D2$ and $\bar D 2$ branes.
In turn, from (\ref{cond2}), the second family presents $D0$ and $D4$ (or $\bar D4$) charges, leading to the interpretation of these systems as
bound states of $D 4$, $\bar D 4$ and $D0$ branes
\footnote{
The solution with $\bar D 0$ charge can be obtained by imposing $\Gamma_{\ob}\epsilon =-\epsilon$ in (\ref{cond2});
the presence of $D4$ charges in the hyper-volumes $(1234),(3456),(5612)$, follows from,
$\Gamma_{1234}= \Gamma_{11}\,\Gamma_{01234}\,\Gamma_{\ob}\sim \Gamma_{11}\,\Gamma_{01234}\,,$ etc.
}.
In the weak coupling analysis, we have constructed the lowest levels of the spectra, and verified the stability (absence of tachyons) of
the vacuum, if the conditions alluded before are satisfied.
Furthermore, each level presents equal number of (space-time) bosons and fermions, which gives strong evidence for SUSY.
This was confirmed by founding the no-force conditions from the computation of the one loop amplitude, and verified that they include the
conditions obtained before, which shows that SUSY holds (at least in this limit) not just for low energies, but to all orders in $\alpha'$.

We would like to remark that, in our analysis, we have assumed that both the brane and the antibrane can be treated as probe charges,
and therefore that they do not affect significatively the space-time, neither does the presence of one brane affect the analysis of the other one.
Even though we have given strong evidences that the brane-antibrane systems could exists, we think that a very interesting open problem
would be the obtention of the long distance solution of SUGRA IIA that describes such systems (being more ambitious, as a limit of some higher
dimensional supertube).
Maybe the very recent results of reference \cite{Denef:2007yt}, where $D6 -\bar D 6$ systems with various $D0-D2-D4$ charges are identified
with black hole solutions of the four dimensional SUGRA obtained after compactification of type II theories on Calabi-Yau threefolds,
points out in the right direction.
\bigskip

\noindent{\bf Acknowledgements.}

We would like to thank Guillermo Silva and Nicol\'as Grandi for discussions, and Mariel Sant\'angelo for reading the manuscript.

\end{document}